\documentclass[11pt,a4paper]{article}
\usepackage[utf8]{inputenc}
\usepackage[T1]{fontenc}
\usepackage[english]{babel}
\usepackage{amsmath,amssymb,amsthm}
\usepackage{booktabs}
\usepackage[margin=2.7cm]{geometry}
\usepackage{xcolor}
\usepackage[colorlinks=true,linkcolor=blue,citecolor=blue,urlcolor=blue]{hyperref}
\usepackage[linesnumbered,ruled,vlined]{algorithm2e}

\newtheorem{theo}{Theorem}[section]
\newtheorem{lemm}[theo]%
 {Lemma}
{Definition}
\newtheorem{prop}[theo]%
{Proposition}
{Corollary}
{Definition-Proposition}
{Conjecture} 
{Remark} 
{Example} 

\newcommand{\finpreuve}{\mbox{} \hfill \mbox{$\Box$}}
\newenvironment{preuve}{\noindent {\it Proof: }}{\finpreuve}

\def\F{{\mathbb F}}

\def\cF{{\mathcal F}}

\newcommand{\ds}{\displaystyle}
\newcommand{\CQ}{\mathcal{C}_{Q}}
\newcommand{\CN}{\mathcal{C}_{N}}

\newcommand{\1}{\mathbf{1}}
\newcommand{\wh}[1]{\widehat{#1}}

\newcommand{\Tr}{\operatorname{Tr}}

\begin{document}

\title{A polynomial-time decoding algorithm\\ up to half the square-root bound\\ for quadratic residue and generalized quadratic residue codes}
\author{Philippe Gaborit\thanks{XLIM (UMR CNRS 7252), Universit\'e de Limoges, 123 avenue Albert Thomas, 87060 Limoges Cedex, France. \texttt{gaborit@unilim.fr}}
\and Philippe Lebacque\thanks{XLIM (UMR CNRS 7252), Universit\'e de Limoges, 123 avenue Albert Thomas, 87060 Limoges Cedex, France. \texttt{philippe.lebacque@unilim.fr}}}
\date{\today}
\maketitle

\begin{abstract}
Binary quadratic residue (QR) codes, introduced by Prange in 1958, are cyclic codes of prime length $p\equiv\pm1\pmod 8$ and dimension $(p+1)/2$ whose minimum distance satisfies the square-root bound $d\ge\sqrt p$. They include the Hamming code $[7,4,3]$ and the Golay code $[23,12,7]$, and were generalized by van Lint and MacWilliams to generalized quadratic residue (GQR) codes of prime-power length $q$, for which the bound $d\ge\sqrt q$ still holds and is attained when $q$ is a square. In coding theory, one usually first finds a lower bound on the minimum distance of a family of codes, and then a polynomial-time algorithm decoding up to half of this bound, as for BCH, Reed--Muller or Gabidulin codes. Until now, GQR codes had resisted this approach: no such algorithm was known, apart from individual short lengths. We give a deterministic algorithm which, for every odd prime power $q\equiv\pm1\pmod 8$, corrects every error pattern of weight at most $\lfloor(\sqrt q-1)/2\rfloor$ in the binary GQR code of length $q$, in $O(q^{9/2})$ operations in $\F_q$. We start from a description of QR codes due to Helleseth, which we extend to GQR codes, and in which the syndrome gives the values of the quadratic character of the error-locator polynomial. We then recover this polynomial by adapting to GQR codes the tools that Kopparty developed for recovering polynomials from noisy character values, with the decoding of dual BCH codes as an application. Since the square-root bound is attained when $q$ is a square, this radius is the largest that can be guaranteed uniformly on the class of GQR codes.
\end{abstract}

\section{Introduction}

\subsection{Quadratic residue codes}

Quadratic residue (QR) codes were introduced by Prange in 1958~\cite{Pra58} and form one of the oldest families of algebraic error-correcting codes. The idea of the construction is elementary. Take a prime $p$, index the $p$ coordinates of a binary word by the elements of $\F_p$, and split the nonzero elements of $\F_p$ into the set $Q$ of squares and the set $N$ of non-squares. A binary cyclic code of length $p$ is determined by the set of exponents $a$ such that $\zeta^a$ is a root of all its codewords, where $\zeta$ is a primitive $p$-th root of unity; the QR code is the cyclic code for which this set is $Q$. The construction makes sense as soon as $2$ is a square modulo $p$, that is for $p\equiv\pm1\pmod 8$, and it gives a code of dimension $(p+1)/2$, hence of rate just above $1/2$. The precise definition is recalled in Section~\ref{sec:prelim}.

Van Lint and MacWilliams~\cite{vLM78} extended the construction from primes to prime powers $q=p^m$: the coordinates are now indexed by the elements of $\F_q$, and the code is defined from the squares of $\F_q^*$. These generalized quadratic residue (GQR) codes exist in the binary case for $q\equiv\pm1\pmod 8$, have dimension $(q+1)/2$, and are invariant under the translations of $\F_q$; they are no longer cyclic when $m\ge2$.

QR codes have been generalized in two other directions, which we only mention. Duadic codes~\cite{LMP84} are the cyclic codes of odd length $n$ obtained by replacing the pair $(Q,N)$ by any partition of the nonzero residues modulo $n$ into two unions of cyclotomic classes exchanged by a multiplication $x\mapsto ax$; they exist when $2$ is a square modulo $n$ and still satisfy a square-root bound. Double circulant quadratic residue codes~\cite{Kar69,Gab02} are the codes of length $2p$ and dimension $p$ with generator matrix $[I\,|\,W]$, where $W$ is the circulant matrix defined by the squares modulo $p$; they exist for every odd prime $p$, in particular for $p\equiv\pm3\pmod 8$ where there is no QR code, and they also satisfy square-root bounds~\cite{Cal83,HV04}. In this paper we only consider QR and GQR codes.

\subsection{Minimum distance of QR and GQR codes}

The minimum distance $d_q$ of a GQR code of length $q$ satisfies the square-root bound
\[
d_q^2\ge q,
\]
proved in~\cite[Ch.~16]{MS77} for $q=p$, where it is refined to $d_p^2-d_p+1\ge p$ when $p\equiv7\pmod 8$, and in~\cite{vLM78} for general $q$. How far this bound is from the truth depends on the case.

When $q$ is a square, the bound is attained: $d_q=\sqrt q$, the minimum being reached by a codeword whose support is the set of roots of $X^{\sqrt q}+X$, a one-dimensional $\F_{\sqrt q}$-subspace of $\F_q$~\cite{vLM78} (see also the final remark of~\cite{Vol}).

When $q=p$ is prime, the bound cannot be attained since $\sqrt p$ is not an integer. The refined bound is attained for $p=7$, but for the next lengths the minimum distance is much larger, for instance $7$ for $p=23$, $11$ for $p=47$ and $17$ for $p=89$, whereas $\sqrt{89}<10$. Asymptotically very little is known. Only minor improvements of the lower bound are known~\cite{MS77,HV04,Vol}, and it is not known whether there is an asymptotically good subfamily of QR codes~\cite{Vol}. For comparison, a random binary linear code of rate $1/2$ and length $n$ has minimum distance close to the Gilbert--Varshamov value $\approx0.11\,n$, that is about $n/9$~\cite{MS77}. In the other direction, Voloch~\cite{Vol} proved that for infinitely many primes $p$ the QR code of length $p$ has minimum distance $O(p/\log\log p)$, and $O(p/\log p)$ under the generalized Riemann hypothesis; he also states, without proof, that there are infinitely many non-square prime powers $q$ for which $d_q=O(q/\log q)$.

The square-root bound is thus the only guarantee available for the whole family of GQR codes, and on the whole family it cannot be improved.

\subsection{A remarkable family of codes}

A bound in $\sqrt q$ may look modest, but QR codes are remarkable in five respects, which we recall in turn: their parameters, their symmetries, their weight enumerators, their connection with number theory and lattices, and their relation with quantum codes.

\emph{Parameters.} At short lengths the parameters of QR codes are excellent: the Hamming code $[7,4,3]$ and the Golay code $[23,12,7]$, which are both perfect, and the $[47,24,11]$ code are QR codes~\cite{MS77}. They remain good well beyond these lengths, and they exist for many lengths for which there is no BCH code. For instance the QR code of length $103$ has parameters $[103,52,19]$, whereas shortening the BCH code $[127,71,19]$ to the same length and the same minimum distance only gives a $[103,47,19]$ code, and the QR code of length $199$ improved the best minimum distance known for its length and dimension~\cite{TTAAJ17}. Beyond length $200$, however, the minimum distances and estimates available~\cite{TTAAJ17} suggest that QR codes fall behind BCH codes of comparable length and rate.

\emph{Symmetries.} Add an overall parity check to the QR code of length $p$, and index the $p+1$ coordinates of the extended code by the projective line $\F_p\cup\{\infty\}$. By the Gleason--Prange theorem, the extended code is invariant under the group $\mathrm{PSL}_2(p)$ acting on the projective line~\cite[Ch.~16]{MS77}. This group is much larger than the group of cyclic shifts, and it is doubly transitive, so that all the coordinates of the extended code play the same role. For the lengths $24$ and $48$, the supports of the codewords of any given weight even form $5$-designs~\cite{AM69}.

\emph{Weight enumerators.} For $p\equiv7\pmod 8$ the extended QR code is self-dual and doubly even, that is, all its weights are divisible by $4$~\cite{MS77,RS98}. A doubly-even self-dual code of length $n$ has minimum distance at most $4\lfloor n/24\rfloor+4$, and is called extremal when this value is reached; by Gleason's theorem, the weight enumerator of an extremal code is completely determined by its length~\cite{RS98}. The extended QR codes are extremal for all their lengths up to $48$, namely $8$, $24$, $32$ and $48$, and again for the lengths $80$ and $104$~\cite{RS98}, so that their weight enumerators are known explicitly. The extended QR code of length $48$ is moreover the only extremal doubly-even self-dual code of this length~\cite{HLTP03}.

\emph{Connection with number theory and lattices.} QR codes are arithmetic objects. Their weights are sums of Legendre symbols, which count the points of hyperelliptic curves over $\F_p$, and this is how the Weil bound gives back the square-root bound~\cite{Hel92,HV04}; the asymptotically bad subfamilies of~\cite{Vol} are obtained from the Chebotarev density theorem. These very good codes are also connected to very good lattices: the extended Hamming code $[8,4,4]$ gives the lattice $E_8$, the extended Golay code $[24,12,8]$ gives the Leech lattice, and the extended ternary QR code of length $48$ gives an extremal unimodular lattice in dimension $48$~\cite{CS99}.

\emph{Quantum codes.} Finally, for $p\equiv7\pmod 8$ the QR code of length $p$ contains its dual, which is its subcode of even weight words~\cite{MS77}. The Calderbank--Shor--Steane construction then gives a quantum code encoding one qubit into $p$ qubits with minimum distance $d_p$; for $p=7$ and $p=23$ these are the Steane code $[[7,1,3]]$ and the quantum Golay code $[[23,1,7]]$~\cite{Ste96}.

\subsection{Distance bounds and decoding}

For most classical families of codes, a lower bound on the minimum distance has been followed by a polynomial-time algorithm decoding up to half of this bound. BCH codes were introduced by Hocquenghem~\cite{Hoc59} and by Bose and Ray-Chaudhuri~\cite{BR60} together with the BCH bound, which guarantees a minimum distance at least the designed distance $\delta$; Peterson~\cite{Pet60} and Gorenstein and Zierler~\cite{GZ61} then gave an algebraic decoder correcting $\lfloor(\delta-1)/2\rfloor$ errors, made efficient by the Berlekamp--Massey algorithm~\cite{Ber68,Mas69}. Reed--Muller codes were defined by Muller~\cite{Mul54}, and Reed~\cite{Ree54} gave the same year a majority-logic decoder correcting up to half their minimum distance. In rank metric, Delsarte~\cite{Del78} introduced maximum rank distance codes as spaces of bilinear forms, and Gabidulin~\cite{Gab85} later decoded them up to half their minimum rank distance. Gabidulin's codes are linear over an extension field, a point of view that Delsarte did not adopt, so the two families do not coincide exactly, but the sequence is the same: first a bound, then a decoder.

Until now, QR codes were an exception: no polynomial-time algorithm decoding up to half the square-root bound was known. The existing algebraic decoders treat one length at a time. They recover the syndromes that the code does not provide through Newton's identities and then compute the error-locator polynomial; this was done by Elia~\cite{Eli87} for the Golay code, and by Reed, Truong and their coauthors for the lengths $31$, $41$, $47$, $71$, $79$, $89$ and $97$~\cite{RYT90,RTCY92,HRTC01,CTRCL03,TSSLC08}. Gr\"obner bases can automate the derivation of such formulas~\cite{CRHT94,ABF07}, and permutation decoding uses the large automorphism group for selected parameters~\cite{BBC23}, but none of these methods comes with a complexity bound polynomial in the length. The BCH decoder does not help either: it needs consecutive elements in the defining set, and if the values of the quadratic character are modelled as independent balanced signs, the longest runs of consecutive squares modulo $p$ have length $O(\log p)$, far below the square-root scale; this is only a heuristic, the arithmetic of such runs being itself a difficult problem~\cite{KS15}. Finally, generic decoders such as information-set decoding have a cost of the form $2^{\Theta(\sqrt n)}$ for $\Theta(\sqrt n)$ errors at rate $1/2$~\cite{CTS16}.

There is one more reason to look for a decoder of QR codes. Kudekar, Kumar, Mondelli, Pfister, \c{S}a\c{s}o\u{g}lu and Urbanke~\cite{KKMPSU17} proved that any sequence of binary linear codes of increasing lengths, whose rates converge and whose permutation groups are doubly transitive, achieves capacity on the binary erasure channel under bit-MAP decoding. By the Gleason--Prange theorem this applies to extended QR codes, as noted in~\cite{KKMPSU17}. QR codes are therefore asymptotically optimal for erasures, which are corrected by linear algebra, although nothing comparable is known for their minimum distance or for the correction of errors.

\subsection{Contribution}

We give the first polynomial-time decoding algorithm valid for the whole family of binary QR and GQR codes.

\medskip
\noindent\emph{Main result} (Theorem~\ref{thm:main}). For every odd prime power $q\equiv\pm1\pmod 8$, there is a deterministic algorithm that corrects any $\lfloor(\sqrt q-1)/2\rfloor$ errors in the binary QR code of length $q$ when $q$ is prime, and more generally in the binary GQR code of length $q$, in $O(q^{9/2})$ operations in $\F_q$.
\medskip

The algorithm has two ingredients.

The first one is a description of QR codes due to Helleseth~\cite{Hel92}, different from the classical one, in which the error positions are elements of $\F_p$ and are the roots of a polynomial $f_E$ with coefficients in $\F_p$; in this description, the syndrome of the received word gives the value of the quadratic character of $f_E$ at every point which is not an error position. Helleseth's description is given for prime lengths, and we extend it to all GQR codes. Decoding then amounts to recovering a polynomial of small degree over $\F_q$ from the values of its quadratic character.

The second ingredient is the set of tools recently developed by Kopparty~\cite{Kop} for recovering low-degree polynomials from noisy character values, which he applied in particular to the decoding of dual BCH codes; they combine Hasse derivatives, multiplicity estimates and Stepanov's polynomial method. We do not apply Kopparty's algorithm as it stands: we adapt his method to the case of GQR codes, in which the character values are known exactly outside the roots of $f_E$ and are not known at these roots. Kopparty's algorithm recovers a monic squarefree polynomial of degree at most $(\epsilon/16)\sqrt q$ from the values of its quadratic character, a proportion $1/8-\epsilon$ of which may be wrong~\cite[Theorem~3.1]{Kop}. In our situation the polynomial is split and the only unknown values are at its roots; with parameters chosen for this case we reach the degree $\lfloor(\sqrt q-1)/2\rfloor$ which is needed to decode up to half the square-root bound, and we do not need to factor polynomials. We write a linear system whose unknowns are a polynomial and its Hasse derivatives, we show that a suitable power of $f_E$ is a nonzero solution, and that the error positions can be read on the multiplicities of the roots of any nonzero solution.

This result places GQR codes in the classical situation described above. The algorithm applies to the whole class of GQR codes and decodes up to half the square-root bound; since this bound is attained when $q$ is a square, no larger decoding radius can be guaranteed for the class as a whole, and for $q$ a square the algorithm corrects up to half the true minimum distance. To the best of our knowledge, the existence of a polynomial-time decoding algorithm valid for the whole family had been open since the introduction of QR codes, more than sixty years ago. For QR codes of prime length, where the minimum distance is a priori larger than $\sqrt p$ but no better bound is available, decoding beyond $\sqrt p/2$ errors remains an open problem.

\medskip
\noindent\emph{Organization.} Section~\ref{sec:prelim} recalls the classical definition of QR codes and presents Helleseth's description for prime lengths. Section~\ref{sec:gqr} extends this description to GQR codes and states the main result. Section~\ref{sec:cs} gives the identity between syndromes and character values, Section~\ref{sec:hasse} the lemma on Hasse derivatives and multiplicities, and Section~\ref{sec:algo} the algorithm and its analysis.

\section{Quadratic residue codes: the classical definition and Helleseth's description}\label{sec:prelim}

In this section $p$ is a prime with $p\equiv\pm1\pmod 8$, and $Q$ and $N$ are the sets of nonzero squares and non-squares of $\F_p$. A binary word of length $p$ is seen as a function $f:\F_p\to\F_2$, and is identified with the element $\sum_{x\in\F_p}f(x)X^x$ of the group algebra $\mathcal R=\F_2[\F_p]$, which is here the ring $\F_2[X]/(X^p-1)$. For a subset $E$ of $\F_p$ we denote by $\1_E$ its indicator function.

\subsection{The classical definition}

We first recall how binary cyclic codes are described by cyclotomic classes. Let $\zeta$ be a primitive $p$-th root of unity in $\F_{2^r}$, where $r$ is the order of $2$ in $\F_p^*$. The Fourier transform of $f:\F_p\to\F_2$ is the function $\wh f:\F_p\to\F_{2^r}$ defined by
\[
\wh f(a)=\sum_{x\in\F_p}f(x)\,\zeta^{ax},
\]
so that $\wh f(a)$ is the value at $\zeta^a$ of the polynomial associated with $f$. The cyclotomic class of $a\in\F_p$ is the set $\{a,2a,4a,\dots\}$. Since $f$ is binary, $\wh f(2a)=\wh f(a)^2$, hence the set of zeros of $\wh f$ is a union of cyclotomic classes. A binary cyclic code of length $p$ is defined by a union $T$ of cyclotomic classes, called its defining set: it is the code
\[
\{f:\F_p\to\F_2\ :\ \wh f(a)=0\ \text{ for all } a\in T\},
\]
which has dimension $p-|T|$.

The condition $p\equiv\pm1\pmod 8$ means that $2\in Q$, so that the cyclotomic class of a square is contained in $Q$: the set $Q$ is a union of cyclotomic classes, and so is $N$. The binary QR codes of length $p$ are the cyclic codes with defining sets $Q$ and $N$,
\[
  \CQ=\{f:\ \wh{f}(a)=0\ \ \forall a\in Q\},\qquad
  \CN=\{f:\ \wh{f}(a)=0\ \ \forall a\in N\}.
\]
As $|Q|=|N|=(p-1)/2$, they have dimension $(p+1)/2$, and they are equivalent through the permutation $x\mapsto sx$ of the coordinates for any $s\in N$. For $p=7$ the cyclotomic classes are $\{0\}$, $Q=\{1,2,4\}$ and $N=\{3,5,6\}$, and $\CQ$ is the Hamming code $[7,4,3]$. For $p=23$ the sets $Q$ and $N$ are themselves cyclotomic classes, with $11$ elements, and $\CQ$ is the Golay code $[23,12,7]$.

In this description, the natural syndromes of a received word $y=c+\1_E$, where $c\in\CQ$ and $E$ is the set of error positions, are the values $\wh y(a)=\wh{\1_E}(a)$ for $a\in Q$. They belong to the field $\F_{2^r}$, whose degree $r$ can be as large as $(p-1)/2$, and the error positions appear as exponents of $\zeta$. The defining set $Q$ contains no long sequence of consecutive integers, so the BCH bound and the BCH decoder give very little, and the values $\wh{\1_E}(a)$ for $a\in N$, which would be needed to compute the error-locator polynomial by Newton's identities, are not available. This is the obstacle met by all the decoders of QR codes based on the classical description.

\subsection{Helleseth's description}

Helleseth~\cite{Hel92} proposed to look at QR codes differently: the error positions are kept as elements of $\F_p$, and one uses the arithmetic of $\F_p$ itself instead of that of $\F_{2^r}$. This is the point of view from which we start.

We first need a description of the code that does not involve $\zeta$. Let $e_Q=\sum_{r\in Q}X^r\in\mathcal R$, and let
\[
A_Q=(\1_Q(x-y))_{x,y\in\F_p}
\]
be the matrix of the multiplication by $e_Q$ in $\mathcal R$. Since $2Q=Q$, we have $e_Q^2=e_Q$, hence $A_Q^2=A_Q$. The ideal of $\mathcal R$ generated by $e_Q$ when $p\equiv7\pmod 8$, and by $1+e_Q$ when $p\equiv1\pmod 8$, is a QR code~\cite[Ch.~16]{MS77}, namely $\CQ$ or $\CN$ according to the choice of $\zeta$. As these elements are idempotents, a parity-check matrix of this code is
\[
H_p=\delta I+A_Q,\qquad\text{with }\delta=1\text{ if }p\equiv7\pmod 8\text{ and }\delta=0\text{ if }p\equiv1\pmod 8,
\]
and the syndrome of a word $y=c+\1_E$ is the binary vector $s=H_py=H_p\1_E$, whose coordinates $s_x$ are indexed by $x\in\F_p$.

Let now $E\subset\F_p$ be a set of error positions, of size $d$. We associate with $E$ the polynomial
\[
f_E(X)=\prod_{a\in E}(X-a)\in\F_p[X].
\]
Contrary to the classical error-locator polynomial, $f_E$ has its coefficients in $\F_p$, a field whose size is the length of the code. Let $\chi$ be the Legendre symbol modulo $p$, and let $x\in\F_p\setminus E$. On the one hand, each factor $\chi(x-a)$ of $\chi(f_E(x))$ equals $1$ if $x-a\in Q$ and $-1$ if $x-a\in N$. On the other hand, by definition of $A_Q$, the coordinate $s_x$ is the parity of the number of $a\in E$ such that $x-a\in Q$. Comparing the two gives Helleseth's identity, in the pointwise form which appears in the proof of~\cite[Lemma~2]{HV04},
\begin{equation}\label{eq:hel}
(-1)^{s_x}=(-1)^d\,\chi\bigl(f_E(x)\bigr)\qquad\text{for every }x\in\F_p\setminus E.
\end{equation}
Helleseth used this relation to express the weights of the words $A_Q\1_E$ as sums of Legendre symbols, which gives back the square-root bound through the Weil bound.

For decoding, the consequence is the following: the syndrome gives the values of $\chi\circ f_E$ at all the points of $\F_p$ outside $E$, and decoding amounts to recovering the polynomial $f_E$, which is squarefree, split and of degree $d$, from these values. This is a problem of reconstruction of a polynomial from character values, of the type considered by Kopparty~\cite{Kop}. The description involves neither the extension field $\F_{2^r}$ nor the cyclic structure.

\section{Generalized quadratic residue codes and the main result}\label{sec:gqr}

\subsection{Extension of Helleseth's description to GQR codes}\label{sec:ext}

Helleseth's description is stated for prime lengths and does not cover GQR codes. Voloch states in~\cite{Vol} that Helleseth's lemma has a generalization to the generalized quadratic residue codes of van Lint and MacWilliams, sketched in~\cite{HV04}; however~\cite{HV04} only deals with prime lengths, and to the best of our knowledge this generalization has not been written down. We do it in this section, with the notation of Section~\ref{sec:prelim}: the field $\F_p$ is replaced by $\F_q$, the functions $x\mapsto\zeta^{ax}$ by the additive characters $x\mapsto\zeta^{\Tr(ax)}$ of $\F_q$, and the Legendre symbol by the quadratic character of $\F_q$. For $q=p$ the definitions below are those of Section~\ref{sec:prelim}, and the identity~\eqref{eq:hel} becomes Proposition~\ref{sc}.

Let $q=p^m$ be a prime power, where $p$ is odd and $2$ is a square in $\F_{q},$ which is equivalent to assuming that $q\equiv \pm 1\pmod 8.$ We denote by $Q$ the set of squares of $\F_{q}^*$ and by $N$ the set of non-squares of $\F_{q}^*.$ Let $\Tr$ denote the trace from $\F_{q}$ to $\F_{p}.$

In this paper we investigate the decoding of the binary GQR-codes constructed as follows. Let us consider a primitive $p$-th root of unity $\zeta\in \F_{2^r},$ where $r$ denotes the order of $2$ in $\F_{p}^*,$ and associate with any function $f:\F_{q}\to \F_{2},$  its Fourier transform  $\wh{f}:\F_{q}\to\F_{2^r}$ defined by
\[
  \wh{f}(a)=\sum_{x\in \F_q}f(x)\,\zeta^{\Tr(ax)}.
\]
 Let $\cF$ be the set of functions from $\F_{q}\to \F_{2},$ identified with $\F_{2}^q$ hereafter, and $\mathcal{E}$  those from $\F_{q}\to\F_{2^r}.$ Consider the sub-vector spaces of $\cF$
\[
  \CQ=\{f\in\F_2^{\,q}:\ \wh{f}(a)=0\ \ \forall a\in Q\},\qquad
  \CN=\{f\in\F_2^{\,q}:\ \wh{f}(a)=0\ \ \forall a\in N\},
\]
These two codes are equivalent. To see that, it suffices to consider an element $s\in N$ and to associate with a function $f\in \CQ$ the function $f(s^{-1}x),$ whose Fourier transform is $\wh{f}(sx).$

There is another way to view these codes. We consider elements $e_{Q}$ and $e_{N}$ in the group algebra $\mathcal{R}=\F_{2}[\F_{q}]$ given by $e_{Q}=\sum_{r\in Q}X^r$ and $e_{N}=\sum_{r\in N}X^r,$ and the matrix of the multiplication (convolution) by $e_{Q}:$
$$A_{Q}=(\1_{Q}(x-y))_{x,y\in\F_{q}},$$ where $\1_{Q}(x)=1$ for squares and $0$ otherwise. As $e_{Q}^2=e_{Q},$ we see that $A_{Q}^2=A_{Q}.$   The matrix $A_{Q}$ generates the code $\CQ$ or $\CN$ if $q\equiv7\pmod 8$ and $I+A_{Q}$ generates the code $\CQ$ or $\CN$ if $q\equiv1\pmod 8.$ As $A_{Q}^2=A_{Q},$ the control matrix is $I+A_{Q}$ in the first case and $A_{Q}$ in the second, that we denote by $H_{q}.$ 

\subsection{Main result}\label{sec:main}

From these facts, it is sufficient to treat the case of the code $C=\ker H_{q}$ in order to establish the following theorem.

\begin{theo}\label{thm:main} For any prime power $q\equiv\pm1\pmod 8,$ the code $\CQ$, which is the binary QR code of length $q$ when $q$ is prime and the binary GQR code of length $q$ in general, is a code with parameters 
\[
 \left[q,\frac{q+1}{2},d_q\right],
\]
with $d_q\ge\sqrt q,$ for which there is a deterministic decoding algorithm for correcting any error of weight 
\[
 t(q)=\left\lfloor\frac{\sqrt q-1}{2}\right\rfloor
\] running in $O(q^{9/2})$ operations in $\F_{q}.$
\end{theo}

Before proving the theorem, we note that we present a uniform algorithm for all generalized QR codes. For $q=s^2,$ the minimum distance is indeed $s$ and thus our algorithm decodes up to the error-correcting capability. Decoding QR codes ($q$ prime) beyond this radius remains an open problem. 

We also present the main ideas behind the algorithm which is based on Helleseth observation and the Kopparty algorithm for recovering polynomials from character values. We map a received codeword to its syndrome and try to find an error vector it comes from. To achieve this, we only use the bound $t(q)$ on the weight of its support.

If we build the split separable polynomial whose set of roots matches the support of the error, that we call $E,$ and use a (slightly generalized) observation of Helleseth, we find that a large power of this polynomial is part of a solution of a linear system involving it and its Hasse derivatives. Thus we solve the linear system for $d=t(q)$ and recover the support of the error from a solution. At the end, we know that if the number of errors is below the decoding radius, we will find the right one.

\section{Character-syndrome identity}\label{sec:cs}

The first ingredient of our proof is an identity that links syndrome and character values of separable split polynomials. Let us denote by $\chi$ the quadratic character of $\F_{q}$ and recall that $\chi(x)=x^{\frac{q-1}{2}}$ for any $x\in\F_{q}.$  We use the following notation, for any subset $E$ of $\F_{q}:$

$$f_{E}(X)=\prod_{a\in E}(X-a).$$ Our algorithm will be partly based on the following identities due to Helleseth~\cite{Hel92} (stated in the case of $q=p,$ see also the proof of~\cite[Lemma~2]{HV04}). We are only going to use the first one in our algorithm.
\begin{prop} \label{sc} Let $E\subset\F_{q},$ $d=|E|,$ $s=H_{q}\1_{E},$ $\delta=0$ if $q\equiv1\pmod 8$ and $\delta=1$ if $q\equiv7\pmod 8.$ Then
$$\forall x\notin E,\ (-1)^{s_{x}}=(-1)^{d}\chi(f_{E}(x))\quad\text{ and }\quad \forall x\in E,\  (-1)^{s_{x}}=(-1)^{d+\delta-1}\chi(f'_{E}(x))$$
\end{prop}
\begin{preuve} Let $x\notin E.$ Then $\ds s_{x}=\sum_{y\in \F_{q}}(\delta I+A_{Q})_{x,y}\1_{E}(y)=\sum_{y\in E}\1_Q(x-y).$ Thus $$(-1)^{s_{x}}=\prod_{y\in E}(-1)^{\1_Q(x-y)}=\prod_{y\in E}(-\chi(x-y))=(-1)^d\chi(f_{E}(x)).$$

Now let $x\in E.$ Then $\ds s_{x}=\sum_{y\in \F_{q}}(\delta I+A_{Q})_{x,y}\1_{E}(y)=\delta+\sum_{y\in E\setminus\{x\}}\1_Q(x-y).$ Thus 
\begin{align*}(-1)^{s_{x}}&=(-1)^\delta\prod_{y\in E\setminus\{x\}}(-1)^{\1_Q(x-y)}=(-1)^\delta\prod_{y\in E\setminus\{x\}}(-\chi(x-y))\\
&=(-1)^{\delta+d-1}\chi(f'_{E}(x)).
\end{align*}
\end{preuve}

Note that these equalities give the lower bound $\sqrt{q}$ through an application of the Weil bounds.

 The second ingredient is an application of the Kopparty algorithm for reconstructing a polynomial from character values of evaluation of this polynomial. As we need here a stronger error-free version of this algorithm, let us use directly Kopparty's ideas and recover this algorithm in our setting.

\section{Hasse derivatives and multiplicities}\label{sec:hasse}

Let us start with a technical lemma on Hasse derivatives allowing to deal with multiplicities of roots of polynomials.

Recall that the $j^{\text{th}}$-Hasse derivatives of $P\in\F_q[X]$, are the polynomial $P^{[j]}$ such that $P(X+T)=\sum_{j\ge0}P^{[j]}(X)T^j$.
\begin{lemm}\label{lem:pseudo}
Let $M\ge1$, $K\ge0$ such that $K+M\le q$, and $0\ne W\in\F_q[X]$ with $\deg W<Mq$.
Assume that for any $0\le j<M$, there is a polynomial $P_j\in\F_q[X]$ of degree at most $K$ such that 
$W^{[j]}=P_j\mod \Lambda,$ where $\Lambda=X^q-X.$
Then for every $a\in\F_q$,
\[
\mathrm{mult}_a(W)\pmod q\ \le K+M-1.
\]
where $\mathrm{mult}_{a}$ is the multiplicity of $a$ as a root of $W$ and the integers modulo $q$ are identified with $\{0,\dots,q-1\}.$ 
\end{lemm}

Let us prove this lemma which is about Hasse derivatives, polynomials and pseudopolynomials. It is a version at the points of $\F_q$ of \cite[Lemma 3.3]{Kop}, combined with the observation of \cite[Section 5.1]{Kop} that the conditions on the first $M$ derivatives bound the degrees of the $A_i.$

\begin{preuve}
Let $\Lambda$ be the polynomial $X^q-X.$ The condition on the degree of $W$ allows us to write  $W$ as $W=\sum_{i=0}^{M-1}A_{i}\Lambda^i,$ where $A_{i}$ are polynomials of degree less than $q.$ As $\Lambda(X+T)=T^q-T+X^q-X,$  it is clear that, 
 $\Lambda^{[j]}=\Lambda$ if $j=0,$ $-1$ if $j=1,$ $1$ if $j=q,$ and $0$ otherwise.

 For $k<M(\leq q),$ this computation and the Leibnitz rule on Hasse derivatives leads to 
$$W^{[k]}=\sum_{i=0}^{M-1}(A_{i}\Lambda^i)^{[k]}=\sum_{i=0}^{M-1}(-1)^iA_{i}^{[k-i]}\mod \Lambda$$ (see \cite[Lemma 2.4]{Kop}) with the convention that $A_{i}^{[k-i]}=0$ if $k-i<0.$

As $W^{[j]}=P_j\mod \Lambda,$ with $\deg P_{j}\leq K,$ it follows by induction on $i$ that $\deg A_{i}\leq K,$ as it equals $\mod \Lambda$ a polynomial of degree less or equal to $K$ and is of degree strictly less than $q.$ 

Let $a\in\F_{q}.$ Recall that $\mathrm{mult}_{a}(W)=r$ if $r$ is the largest integer such that for all $0\leq \ell<r,$ $W^{[\ell]}(a)=0.$
We have 
\[
 \sum_{i=0}^{\deg W} W^{[i]}(a) T^i=W(a+T)=\sum_{i=0}^{M-1}A_i(a+T)(T^q-T)^i
        =\sum_{j=0}^{M-1}T^{qj}C_j(T),
\]
where
\[
 C_j(T)=\sum_{i=j}^{M-1}(-1)^{i-j}\binom ij T^{i-j}A_i(a+T),
\ \text{ and }\deg C_j\le K+M-1-j\le q-1-j.
\]

From the last inequality, we see that every term in the double sum has a different degree and that $qj\leq \deg T^{qj}C_j(T)\leq q(j+1)-1-j.$  Therefore, 
if $j_{0}$ is the smallest $j$ such that  $C_j\neq 0,$ we see immediately that  $\mathrm{mult}_{a}(W)=qj_{0}+\mathrm{ord}_{T}C_{j_{0}},$ the latter being an integer smaller or equal to $K+M-1-j_{0}.$ The lemma follows. 
%\sum_{i=0}^{j}\Lambda_i^{[i]}A_i^{[j-i]}(x)=\sum_{i=0}^{j}(-1)^i A_i^{[j-i]}(x).$$
\end{preuve}

\section{The algorithm}\label{sec:algo}

In this section, we explain how to decode a received word. Assume that we received $m=c+e$ where $c$ is a codeword. We compute $s=H_{q}m=H_{q}e$ where $e$ is unknown. Let $E$ be its support; for $T\subset\F_q$ we write $e_T=\1_T.$ Let $d\geq0$ be an integer. It is a parameter of the interpolation below: it is fixed once and for all, and it will be taken equal to $t(q)$ in the algorithm. We can compute the quantity 
$$r_{d}(x)=(-1)^{s_{x}+d}.$$
If $|E|=e,$ we have
$\chi(f_{E}(x))=(-1)^{d-e}r_{d}(x)$ for any $x\notin E,$ $\chi(f'_{E}(x))=(-1)^{d-e+\delta-1}r_{d}(x)$  and $\chi(f_{E}(x))=0$ for any $x\in E$ (see Proposition \ref{sc}). 
 We know that this system admits at least a solution with a split polynomial of degree at most $t(q)$ (by hypothesis). But the $\chi(f_{E}(x))$ are values of a polynomial of very large degree (it may be about $q^{3/2}/4)$, so that recovering $f_{E}$ from this system seems hopeless. This is why we use the ideas of Kopparty and build an other system taking into account the multiplicities of roots of $f_{E}^{\frac{q-1}{2}}.$
 
Let $h=\frac{q-1}{2}$ so that for any $x\in \F_{q},$ and any polynomial $P\in\F_{q}[X],$ $P^h(x)=\chi(P(x)).$   Let $d\ge1$, and put
 $$M=d+1, \quad u=d^2+d,\quad D=dh+d^2+d,
 \qquad r_d(x)=(-1)^{d+s_x}.$$

Consider the $\F_{q}$ linear system with polynomial unknown $(F,U_0,\ldots,U_{M-1})$ (it is the system of \cite[Algorithm A]{Kop}, with parameters adapted to our situation)
\begin{equation}\label{eq:interpol}
 \left\{\begin{aligned}
 &F^{[j]}(x)=r_d(x)U_j(x) &&(\forall x\in\F_q,\ \forall 0\le j<M),\\
 &\deg F\le D,&&\\
 &\deg U_j\le u &&(\forall 0\le j<M).
 \end{aligned}\right.
\end{equation}

\begin{prop} \label{lem:witness}
If $s=H_qe_E$ with $|E|=e\leq d$, the system ~\eqref{eq:interpol} admits
\[
 F=f_E^{h+M},\qquad U_j=(-1)^{d-e}F^{[j]}/f_E^h
\] as a non zero solution, and $\deg U_j\le eM-j$.
For any $s\in\F_2^{\,q},$ if $u<q$, every non zero solution of ~\eqref{eq:interpol} satisfies $F\ne0$.
\end{prop}
\begin{preuve} Let $j<M.$ From the Leibnitz rule for Hasse derivatives, it is clear that $f_E^{h+M-j}\mid F^{[j]},$ and therefore $U_{j}$ are polynomials vanishing on $E$ and satisfying $\deg U_{j}\leq e(h+M)-j-eh=eM-j\leq u.$ Moreover $\deg F=e(h+M)\leq D.$ If $x\notin E,$ then $r_d(x)=(-1)^{d-e}\chi(f_E(x))=(-1)^{d-e}f_E(x)^h$ by Proposition \ref{sc}, so that $F^{[j]}(x)=(-1)^{d-e}f_{E}(x)^{h}U_{j}(x)=r_d(x)U_j(x),$ and if $x\in E,$ then $F^{[j]}(x)=0=r_{d}(x)U_{j}(x)$. If $F$ is zero and $u<q,$ the polynomials $U_{j}$ have to be zero as they vanish on $\F_{q}$ and are of degree $\leq u<q.$
\end{preuve}

Now that we know that the system has a solution, let us show how to get $E$ out of a solution. The argument follows the alternate analysis of \cite[Section 5.6]{Kop}: the product of two solutions behaves like a polynomial of low degree on $\F_q,$ together with its derivatives. We apply this to the product of an arbitrary solution by the particular one, and to the square of an arbitrary solution.

\begin{prop}\label{prop:extract}
Assume $d\geq1$ and $q\ge(2d+1)^2.$ Let $s\in\F_2^{\,q}$ and let $(G,V_{0},\dots,V_{M-1})$ be a non zero solution of \eqref{eq:interpol}. Let 
\[
 S(G)=\{a\in\F_q:\mathrm{mult}_a(G)\ge q-h-M\}.
\] Then 
\begin{enumerate}
\item[\textup{(i)}] $|S(G)|\le d;$
\item[\textup{(ii)}] if $s=H_qe_E$ with $|E|\leq d$, then $E\subseteq S(G).$
\end{enumerate}
In particular, if $s=H_qe_E$ with $|E|=d$, then $S(G)=E$.
\end{prop}

\begin{preuve} Note that $u=d^2+d<q,$ so that $G\neq0$ by Proposition \ref{lem:witness}. Let $(G^*,V^*_{0},\dots,V^*_{M-1})$ be a second non zero solution of \eqref{eq:interpol}, for the same $s,$ and let $W=GG^*.$ We wish to apply Lemma \ref{lem:pseudo} to $W,$ with $K=2u.$ We have $M+K=2d(d+1)+d+1\leq h+M\leq \frac{1}{2}(\sqrt{q}+q+1)\leq q-1$ (as $q\geq 7),$ and $\deg W\leq 2D<Mq$ (as $\deg G\leq D,$ $\deg G^*\leq D$ and $Mq-2D=q-2d^2-d\geq 2d^2+3d+1>0).$ We have to check that for any $0\le j<M$, there is a polynomial $P_j\in\F_q[X]$ of degree at most $2u$ such that 
$W^{[j]}=P_j\mod \Lambda.$ By the Leibnitz rule for Hasse derivatives, we get 
$$\forall x\in \F_{q},\quad W^{[j]}(x)=\sum_{i=0}^jG^{[i]}(x){G^*}^{[j-i]}(x)=\sum_{i=0}^jr_{d}(x)^2V_{i}(x)V^*_{j-i}(x).$$ Thus $$W^{[j]}=\sum_{i=0}^jV_{i}V^*_{j-i}\mod \Lambda,$$ where $\deg V_{i}V^*_{j-i}\leq 2u=K.$
The lemma leads us to \begin{equation}\label{eq:multprod}\forall a\in \F_{q}\quad \mathrm{mult}_{a}(GG^*)\bmod q \leq K+M-1< h+M.\end{equation}

\emph{Proof of} (i). We apply \eqref{eq:multprod} with $G^*=G.$ Let $a\in S(G)$ and $m=\mathrm{mult}_a(G),$ so that $m\geq q-h-M=h-d.$ If $m\leq h,$ then $2h-2d\leq 2m\leq 2h=q-1,$ hence $\mathrm{mult}_a(G^2)\bmod q=2m\geq 2h-2d.$ As $h\geq 2d^2+2d>3d,$ we have $2h-2d>h+d=h+M-1,$ which contradicts \eqref{eq:multprod}. Therefore $m\geq h+1$ for every $a\in S(G),$ and 
\[
(h+1)|S(G)|\leq \deg G\leq D=d(h+1)+d^2.
\]
As $d^2<h+1,$ we get $|S(G)|\leq d.$

\emph{Proof of} (ii). Let $|E|=e\leq d.$ By Proposition \ref{lem:witness}, we can take for $G^*$ the solution $F=f_{E}^{h+M},$ with $U_{j}=(-1)^{d-e} F^{[j]}/f_{E}^h.$ Let $a\in E.$ We have $\mathrm{mult}_{a}(GF)=h+M+\mathrm{mult}_{a}(G).$ If $\mathrm{mult}_{a}(G)<q-h-M,$ then $h+M\leq\mathrm{mult}_{a}(GF)<q,$ which contradicts \eqref{eq:multprod}. Thus $\mathrm{mult}_{a}(G)\geq q-h-M$ for any $a\in E$ and therefore $E\subseteq S(G).$

Finally, if $|E|=d,$ then $E\subseteq S(G)$ and $|S(G)|\leq d,$ so that $S(G)=E.$
\end{preuve}

Proposition \ref{prop:extract} has the following consequence, on which the algorithm relies: a single interpolation, with $d=t(q),$ deals with all the error weights $e\leq t(q).$ Whatever the weight $e\leq d$ of the error, any non zero solution $G$ of \eqref{eq:interpol} gives a set $S(G)$ of at most $d$ positions which contains $E,$ and the error is then determined by the syndrome.

We can now state our algorithm and compute its complexity.

\begin{algorithm}[H]
\caption{Decoding algorithm for an entry $m$}
\SetKwInOut{Input}{Input}
\SetKwInOut{Output}{Output}

\Input{$m$, $\F_q$}
\Output{error $e$ }

$t \leftarrow \lfloor \frac{\sqrt{q} - 1}{2} \rfloor$\;
$s \leftarrow H_q(m)$\;

\If{$s = 0$}{
    \Return $e = 0$\;
}

$d \leftarrow t$\;
Compute $r_d,$  the parameters $h, M, u, D$ and solve \eqref{eq:interpol}.

\If{a non-zero solution $F$ exists}{
    For any $a\in\F_{q},$ compute $\mathrm{mult}_{a}(F)$ and deduce $S(F);$

    Solve the system $H_q e = s$ with $\mathrm{Supp}(e)\subseteq S(F)$ (linear system over $\F_2$)\;
    \If{there is a solution $e$}{
        \Return $e$\;
    }
}
\Return \texttt{"Error is outside of our decoding radius"}\;

\end{algorithm}

\medskip

Let us analyze why the algorithm returns the error if $|E|\leq t$ and give an upper bound on the required number of operations. From the choice of our parameters, the system \eqref{eq:interpol} is a system with $q(d+1)$ equations and $d\frac{q-1}{2}+d^2+d+1+(d+1)(d^2+d+1)$ unknowns. We solve it by dense Gaussian elimination, without using the structure of the system, in $O(q^2d^2(d^3+dq))$ operations which is here $O(q^{\frac{9}{2}})$ operations as $d\ll\sqrt{q}.$ It is solved only once. Finding roots and multiplicities of a polynomial of degree $O(q^{\frac{3}{2}})$ over $\F_{q}$ costs $O(q^3)$ operations by the naive Horner method. Finally, the last step is a linear system over $\F_2$ with $q$ equations and $|S(F)|\leq t$ unknowns (Proposition \ref{prop:extract}), so the overall cost of our algorithm is indeed $O(q^{9/2}).$

Now, if there are at most $t(q)$ errors, Proposition \ref{lem:witness} applied with $d=t$ shows that a non zero solution of the system \eqref{eq:interpol} exists, and Proposition \ref{prop:extract} that any non zero solution $F$ satisfies $E\subseteq S(F)$: the vector $e_E$ is a solution of the system of the last step. It is the only one. Indeed, $|S(F)|\leq t$ by Proposition \ref{prop:extract}, so that every solution of this system has weight at most $t$; two solutions differ by a codeword of weight at most $t<\sqrt q,$ which is zero as the minimum distance is $\geq\sqrt q.$ The algorithm therefore returns the error. (When $t=0,$ that is $q=7,$ only the case $s=0$ is relevant.)

The bound $|S(F)|\leq t$ does not depend on the syndrome. Hence, when there are more than $t(q)$ errors, the algorithm either reports a failure or returns a vector $e$ of weight at most $t$ with $H_qe=s$: then $m+e$ is the unique codeword at distance at most $t$ from $m.$

\section*{Acknowledgment}

AI tools (Claude, Anthropic, and ChatGPT, OpenAI) were used during the preparation of this manuscript for language editing, \LaTeX{} assistance, and exploratory discussions. All mathematical statements, proofs, and conclusions were independently verified by the authors, who take full responsibility for the content.


\begin{thebibliography}{99}

\bibitem{AM69} E.~F.~Assmus and H.~F.~Mattson, New $5$-designs, \emph{J. Combin. Theory} 6 (1969), 122--151.

\bibitem{ABF07} D.~Augot, M.~Bardet and J.-C.~Faug\`ere, On formulas for decoding binary cyclic codes, in \emph{Proc. IEEE Int. Symp. Inform. Theory (ISIT 2007)}, 2646--2650.

\bibitem{Ber68} E.~R.~Berlekamp, \emph{Algebraic Coding Theory}, McGraw-Hill, New York, 1968.

\bibitem{BR60} R.~C.~Bose and D.~K.~Ray-Chaudhuri, On a class of error correcting binary group codes, \emph{Inform. Control} 3 (1960), 68--79.

\bibitem{BBC23} H.~Boualame, M.~Belkasmi and I.~Chana, Efficient decoding algorithm for binary quadratic residue codes using reduced permutation sets, \emph{J. Comput. Sci.} 19(4) (2023), 526--539.

\bibitem{Cal83} R.~Calderbank, A square root bound on the minimum weight in quasi-cyclic codes, \emph{IEEE Trans. Inform. Theory} 29(3) (1983), 332--337.

\bibitem{CTS16} R.~Canto Torres and N.~Sendrier, Analysis of information set decoding for a sub-linear error weight, in \emph{Post-Quantum Cryptography (PQCrypto 2016)}, Lecture Notes in Comput. Sci. 9606, Springer, 2016, 144--161.

\bibitem{CTRCL03} Y.~Chang, T.-K.~Truong, I.~S.~Reed, H.~Y.~Cheng and C.~D.~Lee, Algebraic decoding of $(71,36,11)$, $(79,40,15)$, and $(97,49,15)$ quadratic residue codes, \emph{IEEE Trans. Commun.} 51(9) (2003), 1463--1473.

\bibitem{CRHT94} X.~Chen, I.~S.~Reed, T.~Helleseth and T.-K.~Truong, Use of Gr\"obner bases to decode binary cyclic codes up to the true minimum distance, \emph{IEEE Trans. Inform. Theory} 40(5) (1994), 1654--1661.

\bibitem{CS99} J.~H.~Conway and N.~J.~A.~Sloane, \emph{Sphere Packings, Lattices and Groups}, 3rd ed., Springer, New York, 1999.

\bibitem{Del78} P.~Delsarte, Bilinear forms over a finite field, with applications to coding theory, \emph{J. Combin. Theory Ser. A} 25 (1978), 226--241.

\bibitem{Eli87} M.~Elia, Algebraic decoding of the $(23,12,7)$ Golay code, \emph{IEEE Trans. Inform. Theory} 33(1) (1987), 150--151.

\bibitem{Gab85} E.~M.~Gabidulin, Theory of codes with maximum rank distance, \emph{Problems Inform. Transmission} 21(1) (1985), 1--12.

\bibitem{Gab02} P.~Gaborit, Quadratic double circulant codes over fields, \emph{J. Combin. Theory Ser. A} 97 (2002), 85--107.

\bibitem{GZ61} D.~Gorenstein and N.~Zierler, A class of error-correcting codes in $p^m$ symbols, \emph{J. Soc. Indust. Appl. Math.} 9 (1961), 207--214.

\bibitem{HRTC01} R.~He, I.~S.~Reed, T.-K.~Truong and X.~Chen, Decoding the $(47,24,11)$ quadratic residue code, \emph{IEEE Trans. Inform. Theory} 47(3) (2001), 1181--1186.

\bibitem{Hel92} T.~Helleseth, Legendre sums and codes related to QR codes, \emph{Discrete Appl. Math.} 35(2) (1992), 107--113.

\bibitem{HV04} T.~Helleseth and J.~F.~Voloch, Double circulant quadratic residue codes, \emph{IEEE Trans. Inform. Theory} 50(9) (2004), 2154--2155.

\bibitem{Hoc59} A.~Hocquenghem, Codes correcteurs d'erreurs, \emph{Chiffres} 2 (1959), 147--156.

\bibitem{HLTP03} S.~K.~Houghten, C.~W.~H.~Lam, L.~H.~Thiel and J.~A.~Parker, The extended quadratic residue code is the only $(48,24,12)$ self-dual doubly-even code, \emph{IEEE Trans. Inform. Theory} 49(1) (2003), 53--59.

\bibitem{Kar69} M.~Karlin, New binary coding results by circulants, \emph{IEEE Trans. Inform. Theory} 15(1) (1969), 81--92.

\bibitem{KS15} S.~V.~Konyagin and I.~E.~Shparlinski, Quadratic non-residues in short intervals, \emph{Proc. Amer. Math. Soc.} 143(10) (2015), 4261--4269.

\bibitem{Kop} S.~Kopparty, Recovering polynomials over finite fields from noisy character values, arXiv:2601.07137 (2026); to appear in \emph{Proc. FOCS 2026}.

\bibitem{KKMPSU17} S.~Kudekar, S.~Kumar, M.~Mondelli, H.~D.~Pfister, E.~\c{S}a\c{s}o\u{g}lu and R.~L.~Urbanke, Reed--Muller codes achieve capacity on erasure channels, \emph{IEEE Trans. Inform. Theory} 63(7) (2017), 4298--4316.

\bibitem{LMP84} J.~S.~Leon, J.~M.~Masley and V.~Pless, Duadic codes, \emph{IEEE Trans. Inform. Theory} 30(5) (1984), 709--714.

\bibitem{MS77} F.~J.~MacWilliams and N.~J.~A.~Sloane, \emph{The Theory of Error-Correcting Codes}, North-Holland Mathematical Library 16, North-Holland, Amsterdam, 1977.

\bibitem{Mas69} J.~L.~Massey, Shift-register synthesis and BCH decoding, \emph{IEEE Trans. Inform. Theory} 15(1) (1969), 122--127.

\bibitem{Mul54} D.~E.~Muller, Application of Boolean algebra to switching circuit design and to error detection, \emph{IRE Trans. Electron. Comput.} EC-3 (1954), 6--12.

\bibitem{Pet60} W.~W.~Peterson, Encoding and error-correction procedures for the Bose--Chaudhuri codes, \emph{IRE Trans. Inform. Theory} 6(4) (1960), 459--470.

\bibitem{Pra58} E.~Prange, Some cyclic error-correcting codes with simple decoding algorithms, Technical Note AFCRC-TN-58-156, Air Force Cambridge Research Center, Bedford, MA, April 1958.

\bibitem{RS98} E.~M.~Rains and N.~J.~A.~Sloane, Self-dual codes, in V.~S.~Pless and W.~C.~Huffman (eds.), \emph{Handbook of Coding Theory}, vol.~1, Elsevier, Amsterdam, 1998, 177--294.

\bibitem{Ree54} I.~S.~Reed, A class of multiple-error-correcting codes and the decoding scheme, \emph{Trans. IRE Prof. Group Inform. Theory} 4 (1954), 38--49.

\bibitem{RTCY92} I.~S.~Reed, T.-K.~Truong, X.~Chen and X.~Yin, The algebraic decoding of the $(41,21,9)$ quadratic residue code, \emph{IEEE Trans. Inform. Theory} 38(3) (1992), 974--985.

\bibitem{RYT90} I.~S.~Reed, X.~Yin and T.-K.~Truong, Algebraic decoding of the $(32,16,8)$ quadratic residue code, \emph{IEEE Trans. Inform. Theory} 36(4) (1990), 876--880.

\bibitem{Ste96} A.~M.~Steane, Simple quantum error-correcting codes, \emph{Phys. Rev. A} 54 (1996), 4741--4751.

\bibitem{TTAAJ17} M.~Tomlinson, C.~J.~Tjhai, M.~A.~Ambroze, M.~Ahmed and M.~Jibril, \emph{Error-Correction Coding and Decoding: Bounds, Codes, Decoders, Analysis and Applications}, Springer, 2017.

\bibitem{TSSLC08} T.-K.~Truong, P.-Y.~Shih, W.-K.~Su, C.-D.~Lee and Y.~Chang, Algebraic decoding of the $(89,45,17)$ quadratic residue code, \emph{IEEE Trans. Inform. Theory} 54(11) (2008), 5005--5011.

\bibitem{vLM78} J.~H.~van Lint and F.~J.~MacWilliams, Generalized quadratic residue codes, \emph{IEEE Trans. Inform. Theory} 24(6) (1978), 730--737.

\bibitem{Vol} J.~F.~Voloch, Asymptotics of the minimal distance of quadratic residue codes, preprint, \url{https://web.ma.utexas.edu/users/voloch/Preprints/asyquad2.pdf}; also in \emph{Oberwolfach Reports}.

\end{thebibliography}
\end{document}